# Ending AIDS in South Africa: How long will it take? How much will it cost?

Brian G. Williams,† Eleanor Gouws*

† South African Centre for Epidemiological Modelling and Analysis (SACEMA), Stellenbosch, South Africa
* UNAIDS Regional Office for Eastern and Southern African, Johannesburg, South Africa

Correspondence to BrianGerardWilliams@gmail.com

**Abstract**

South Africa has more people infected with HIV but, by providing access to anti-retroviral therapy (ART), has kept more people alive than any other country. The effectiveness, availability and affordability of potent anti-retroviral therapy (ART) make it possible to contemplate ending the epidemic of HIV/AIDS. We consider what would have happened without ART, the impact of the current roll-out of ART, what might be possible if early treatment becomes available to all, and what could have happened if ART had been provided much earlier in the epidemic.

In 2013 the provision of ART has reduced the prevalence of HIV from an estimated 15% to 9% among adults not on ART, the annual incidence from 2% to 0.9%, and the AIDS related deaths from 0.9% to 0.3% *p.a.* saving 1.5 million lives and US$727M.

Regular testing and universal access to ART could reduce the prevalence among adults not on ART in 2023 to 0.06%, annual incidence to 0.05%, and eliminate AIDS deaths. Cumulative costs between 2013 ands 2023 would increase by US$692M only 4% of the total cost of US$17Bn.

If a universal testing and early treatment had started in 1998 the prevalence of HIV among adults not on ART in 2013 would have fallen to 0.03%, annual incidence to 0.03%, and saved 2.5 million lives. The cost up to 2013 would have increased by US$18Bn but this would have been cost effective at US$7,200 per life saved.

Future surveys of HIV among women attending ante-natal clinics should include testing women for the presence of anti-retroviral drugs, measuring their viral loads, and using appropriate assays for estimating HIV incidence. These data would make it possible to develop better and more reliable estimates of the current state of the epidemic, the success of the current ART programme, levels of viral load suppression for those on ART and the incidence of infection.

## Introduction

The prevalence of HIV infection among women attending ante-natal clinics (ANC) in South Africa has increased logistically over time and in all provinces appears to have reached a steady state (Appendix 1). In the case of the Free State the prevalence of HIV infection has not changed for ten years. However, South Africa now has more people receiving anti-retroviral therapy than any other country in the world and this has undoubtedly had a significant impact on the epidemic. If the South African Government follows the current recommendations of the World Health Organization (WHO) and raises the CD4$^+$ threshold for starting treatment from 350 to 500 cells/μL[1] it will be important to estimate the further impact that this will have on the epidemic of HIV and how much it will cost. For completeness we also consider what would have happened if the South African government had implemented the treatment recommendations made by the International AIDS Society (IAS) in 1996[2,3] and the Department of Health and Human Services of the United States of America (DHHS) in 2000.[3,4]

In this paper we use trend data for the prevalence of HIV among women attending ante-natal clinics in South Africa and the reported coverage of anti-retroviral therapy. We fit these data to a dynamical model to estimate current, and to project future, trends in prevalence, incidence, treatment needs and deaths. We estimate the *per capita* cost of HIV/AIDS to the country including the cost of providing drugs and providing support to people on ART, and the cost of hospitalization and access to primary health care facilities. We do not include the social and economic costs incurred when people die of AIDS so that this calculation is conservative with regards to the overall cost to society.

## Data

The model is fitted to data from the annual national surveys of HIV among women attending ante-natal clinics (ANCs).[5-12] The prevalence of HIV is generally higher in pregnant women than in women in the general population and is higher in women than in men. The prevalence of infection among ANC women has therefore been scaled down to match the prevalence in all adults following UNAIDS.[13]

Table 1. Cost of days in hospital, primary health care visits and counselling and testing from Granich *et al.*[14] Costs are for the year 2013.

| Item | Cost ($ *p.a.*) |
|---|---|
| Inpatient days not on ART | 568 |
| Inpatient days on ART | 138 |
| Primary health care visits not on ART | 154 |
| Primary health care visits on ART | 269 |
| Counselling and testing per test | 11 |
| Community support and care | 150 |

Costs include hospitalization, primary health care and treatment.[14] The costs are given in Table 1. We do not discount future costs because of the uncertainty in the economic future. However, the data presented here could be used to discount the costs and benefits at any desired rate.

Table 2. Annual drug costs for triple therapy in South Africa (US$).

| Year | Cost (US$) |
|---|---|
| 1995 | 10,000[15] |
| 2003 | 1,700[16] |
| 2006 | 730[17] |
| 2009 | 188[14] |
| 2014 | 100[18,19] |

The cost of combination ART has declined since it first became available as shown in Table 2. The cost data are fitted to a logistic curve in Figure 1. We assume that the cost of ART will converge to an asymptote of US$50 per year. Below that the drug costs are a small fraction of the total costs of care and support which we set to US$250 per year.

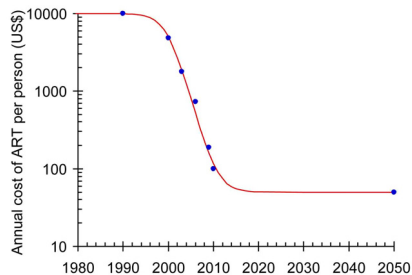

Figure 1. Cost of ART in South Africa (blue dots) fitted to a logistic curve (red line). Data are in Table 1.

## Model

The model is a standard dynamical model discussed in detail elsewhere.[20,21] It includes uninfected people who are susceptible to infection while infected people go through four stages of infection to death to match the known Weibull survival for people infected with HIV but not ART.[22] To account for heterogeneity in the risk of infection the transmission parameter declines exponentially with the prevalence of HIV.[20,21]

**Fitting the model**

We first fit the model to the prevalence data without including ART to get a null model against which to compare the impact of ART. We vary the prevalence of infection in 1980, which determines the timing of the epidemic, the rate of increase and the rate at which the risk of infection declines as the prevalence increases as this determines the peak value of the prevalence. The fitted values are in Table 3.

To model the current level of ART provision we assume that certain proportions of people in the third and fourth stages of HIV infection are started on ART. We assume that coverage increases logistically and we vary the rate and timing of the increase and the proportion of people starting treatment in each stage to match the reported coverage of ART.[13] To explore the impact of active case finding we assume that coverage increases logistically at a realistic rate, timing and asymptotic coverage. We assume that all those that are found to be HIV-positive, in any stage of infection, are eligible for treatment. The parameter values are given in Table 4.

Table 3. The birth, background mortality and transition rates are fixed;[20] the other parameters are varied to optimize the fit to the trend in the prevalence of HIV shown in Figure 2 (below).

| Parameter | Value |
|---|---|
| Adult population in 2013 (M) | 35.0 |
| Birth rate/yr | 0.029 |
| Background mortality/yr | 0.018 |
| Force of infection/yr | 0.791 |
| Prevalence at which transmission is halved | 0.052 |
| Transition rate/yr between HIV stages off ART | 0.348 |
| Transition rate/yr between HIV stages on ART | 0.087 |

Table 4. Model parameters are for logistic functions. *Coverage* gives the asymptotic coverage; *Rate* the exponential rate of increase; *Half-max* the year when coverage reaches half the maximum value. *OR* is the odds-ratio for the number tested to the prevalence of HIV in that stage.

| | Parameter | Value |
|---|---|---|
| Passive case finding: Stage 4 | Coverage | 0.75 |
| | Rate *per annum* | 0.60 |
| | Half-max. (year) | 2008.4 |
| | OR (testing) | 1 |
| Passive case Finding: Stage 3 | Coverage | 0.35 |
| | Rate *per annum* | 0.81 |
| | Half-max. (year) | 2011.6 |
| | OR (testing) | 2 |
| Active case finding: testing | Coverage | 0.90 |
| | Rate *per annum* | 2.00 |
| | Half-max. (year) | 2014.0 |
| | Test interval (yrs) | 1.00 |
| | OR (testing) | 10 |
| Active case finding: take-up | Coverage | 0.90 |
| | Rate *per annum* | 2.00 |
| | Half-max. (year) | 2014.0 |

The model is set up so that a proportion of HIV-positive people are tested a certain number of times each year and of those that test positive a proportion start ART. However, many people who are not infected will also be tested and in order to determine the cost of testing we need to know the case detection rate, that is the proportion of all those tested that are infected with HIV. Under passive case-finding people present to a health-service in stages 3 or 4 of HIV infection. We assume that all those that present in stage 4 will be infected with HIV but that only 50% of those that present in stage 3 will be infected with HIV. Under active case finding we assume that only 10% of those that are tested will be infected with HIV. Since we also have to avoid the mathematical possibility that we test more people than there are in the population we set the odds ratio for the number tested to the number that are HIV-positive to 1 for stage 4, 2 for stage 3, and 10 for active case-finding; when the prevalence is low the number that are positive is 100%, 50% or 10%, respectively, of the number tested but when the prevalence is high the number tested in each test interval never exceeds the total population.

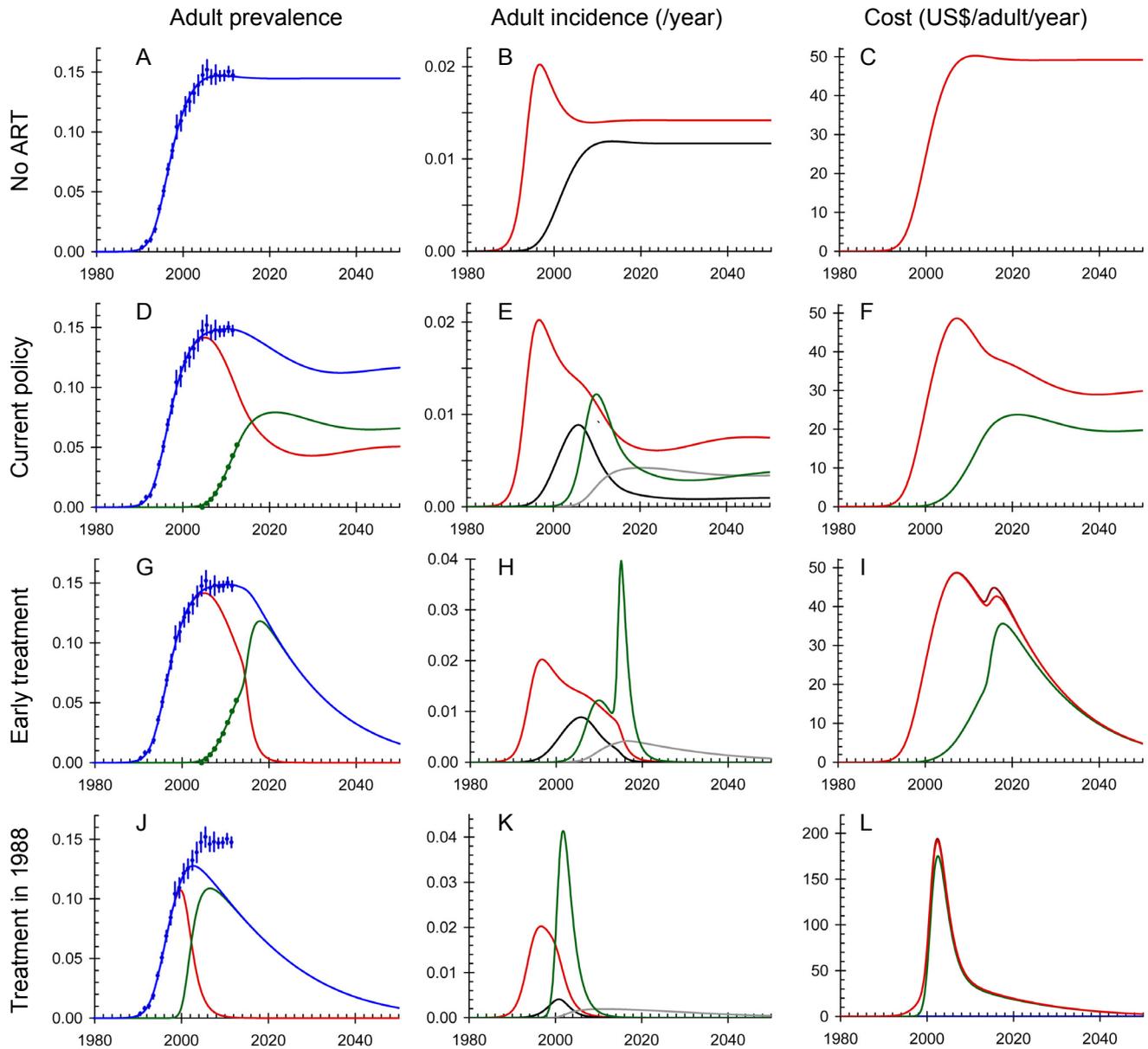

Figure 2. *Top row:* baseline scenario without ART. *Second row:* current level of treatment. *Third row:* early treatment (see text for details). *Bottom row:* Early treatment starting in 1988 (see text for details). *Left:* HIV prevalence. Blue: Data (with 95% confidence limits) and fitted line. Red: not on ART. Green: Data for those on ART and fitted line. *Middle:* HIV incidence. Red: Incidence of HIV. Green: incidence of treatment. Black: mortality not on ART. Grey: mortality on ART. *Right:* Cost. Green: ART. Red: Green plus hospitalization. Brown: Red plus cost of testing.

## Results

The fitted data and implied trends are shown in Figure 2 using the parameter values in Table 3 and Table 4. Figure 2 gives, from left to right, the prevalence of infection, the incidence of infection and of treatment, and the costs. From top to bottom the graphs give the counterfactual that would have happened without ART, the impact of the current level of ART, the predicted impact of universal access to early treatment starting in 2014 and the impact of universal access to early treatment if it had started in 1998 following the 1996 IAS[2,3] and 2000 DHHS[3,4] guidelines.

### Without ART

Figure 2A gives the prevalence of HIV, Figure 2B the implied incidence (red line) and mortality (black line) and Figure 2C the implied costs of in-patient and out-patient care[14] (red line). The prevalence rises rapidly to a steady state and remains fairly constant after 2005. The incidence peaks, declines as people start to die, and then levels off at about 1.4% per year. The mortality rises about ten years after the incidence, reflecting the mean life-expectancy of people with HIV, and levels off at 1.2% per year. The annual cost to the health system would have increased steadily to US$50 per adult per year, or US$1.8Bn per year, in 2013 and then remained constant.

**Progress so far**

The provision of ART on a significant scale in the public sector in South Africa began in 2005 and expanded rapidly. Following the then recommendations of the World Health Organization (WHO)[23] infected people only started ART when their $CD4^+$ cell count fell to 200/μL or they were in clinical stages 3 or 4.[24] In 2006 WHO increased the $CD4^+$ cell count at which people could start ART to 350/μL and the South African government followed suit.

Figure 2D gives the HIV prevalence of those not on ART (red line), the reported number of people on ART (green dots) the fitted number of people on ART (green line) and all infected people (blue line); Figure 2E gives the incidence of HIV (red line), the rate at which people start ART (green line), the mortality of those on ART (black line) and of those not on ART (grey line). Figure 2F gives the cost of ART (green line)[14] and the cost of ART plus the cost of in-patient and out-patient care (red line).

Currently (2013), an estimated 14% of adults are infected with HIV (Figure 2D: blue line), the prevalence of HIV positive adults not on ART (Figure 2D: red line) has fallen to 8% and the prevalence on ART has increased to 6%. If the current policy were to continue the prevalence of HIV positive adults not on ART would continue to fall to about 5%.

There has been a similarly large impact on incidence which has fallen from a peak of 2% *per annum* to 0.9% *per annum* in 2013 (Figure 2E: red line) with the rate at which people start ART peaking at 1.2% *per annum* in 2010 (Figure 2E: green line) but falling rapidly after that. AIDS related deaths among adults peaked in 2006 at about 0.9% *per annum* but by 2013 had fallen to 0.3% *per annum* (Figure 2E: black line). Without ART about 420 thousand people would have died in 2013; the provision of ART has reduced this to 102 thousand saving more than 318 thousand lives in 2013. People will continue to die on ART (Figure 2E: grey line) but mainly from natural causes other than HIV.

While the current ART provision has saved many lives and greatly reduced incidence, there has also been a substantial cost saving—the annual cost in 2013 has fallen from $50 per adult, what it would have been without ART, (Figure 2C: red line) to $41 per adult (Figure 2F: red line) for a net cost saving of about $315M in the year 2013 alone. The total cost in 2013 is US$1.4Bn of which 44% is the cost of providing ART, including care and support, (Figure 2F: green line) and 56% is the cost of in-patient and out-patient care (Figure 2F: difference between red and green lines).

**Universal Access to Early Treatment**

The substantial impact of treatment on the incidence, prevalence and mortality due to AIDS and the savings that accrue, encourages one to consider the impact of making ART available to everyone, regardless of $CD4^+$ cell counts. If the South African government chooses to adopt the 2013 guidelines of the World Health Organization[25] about 90% of all HIV-positive people will be eligible for treatment as soon as they are found to be HIV-positive. The current (2013) guidelines of the International AIDS Society[26] and the Department of Health and Human Services (DHHS)[27] both recommend treatment for those infected with HIV without regard to their $CD4^+$ cell count on the grounds that this is in the best interests of the individuals concerned and has the added benefit of reducing the likelihood that they will infect their partners. We therefore consider what would happen under a policy of universal access in which 90% of all adults, not on ART, are tested each year and 90% of those that test positive are started on ART, with coverage and testing reaching half the target levels by the end of 2014 (Table 4).

Universal access to early treatment should eliminate HIV transmission (Figure 2G and Figure 2H: red lines) and eliminate AIDS related deaths by 2018 (Figure 2H: black line) but there will still be a very large number of HIV positive people on ART (Figure 2G: green line) who will have to be maintained on treatment for the rest of their lives. The rate at which people need to be started on treatment will increase to about three times the present rate (Figure 2H: green line) in 2015 but after that will fall rapidly as transmission and the generation of new cases fell. The costs will begin to fall rapidly (Figure 2I: red and brown lines) as the prevalence of infection falls. There will be a small initial increase in costs as the backlog of untreated patients is taken up. The cumulative costs to 2023 would increase by US$157M. South Africa currently spends about US$500 *per capita* on health so that a policy of early treatment will only increase overall costs by about 0.8% of total spending on health up to 2023 but by then the annual cost will have fallen from $35 per adult, under the current policy to $31, per adult, under the policy of early treatment and after that there would be further cost savings.

**Starting treatment in 1998**

Finally, we make a comparison of what would have happened if a policy of early treatment had been started in 1998 once triple combination therapy became available following the 1996 guidelines of the IAS and the 2000 guidelines of the DHHS.[3] The impact that this policy would have had is shown in the bottom row of Figure 2. By 2010 all HIV positive people would have been on ART (Figure 2J: red line) and incidence would have been eliminated (Figure 2K: red line). A massive campaign of testing and starting people on treatment would have been necessary (Figure 2K: green line) but by 2013 that would have been over. Costs would have been dominated by the high costs of treatment in 2000 (Figure 2L: green line) but by 2013 total costs would have fallen to US$29 per adult per year down from a total cost of US$1.4Bn per year to US$644M per year.

**Bringing it all together**

Figure 3 is a direct comparison of what would have happened without ART, what has happened as a result of the roll-out of ART, what could happen under universal access and early treatment following the new WHO guidelines, and what could have happened if a policy of early treatment had been started in 1998 following the 1996 recommendations of the IAS[2] and the 2000 recommendations of the DHHS.[2,4] The current policy has saved, and will continue to save, both lives (Figure 3A and B) and money (Figure 3C and D). Universal access to early treatment will incur a small increase in costs as treatment is expanded (red and green lines (Figure 3C and D) but will save many more lives (Figure 3A and B) and, in the long run, save more money (Figure 3C and D), eliminate HIV and bring an end to AIDS related deaths (Figure 3A and B).

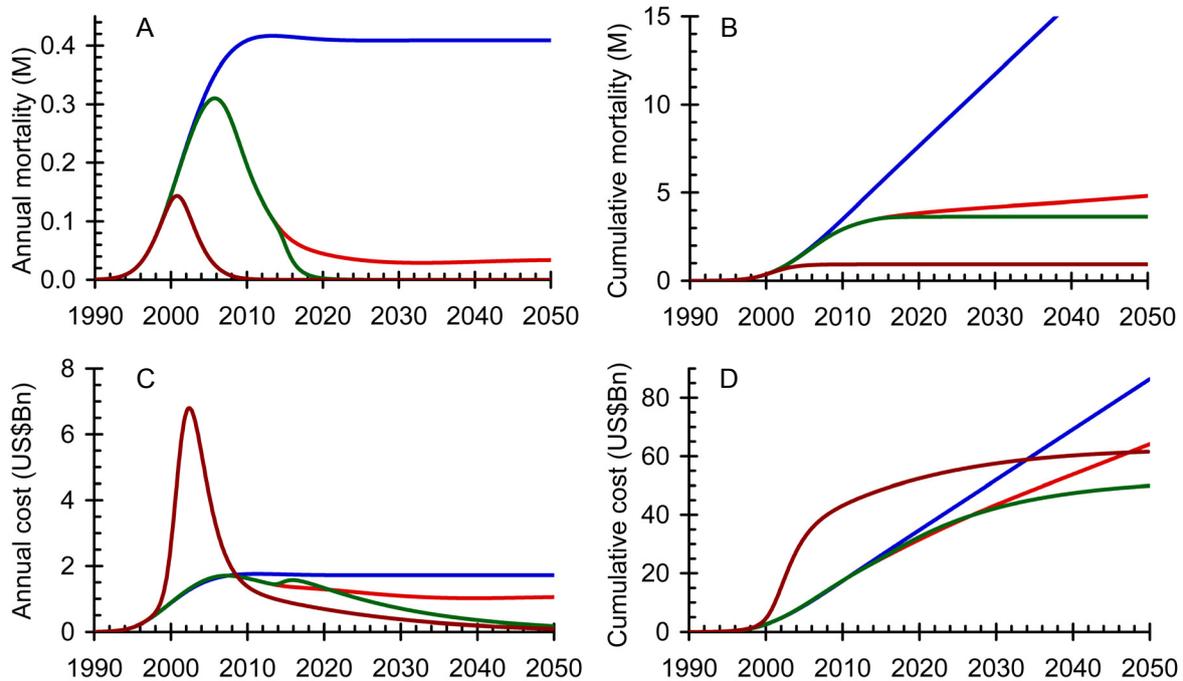

Figure 3. A: Annual mortality from AIDS; B: Cumulative mortality from AIDS; C: Annual cost of ART plus in-patient and out-patient care for AID related conditions; D: Cumulative costs of ART plus in-patient and out-patient care for AID related conditions. Blue line: counterfactual if ART had not been made available; Red line: continuation of current policy on ART provision; Green: provision of early treatment and universal access; Brown line: counterfactual if a policy of early treatment had been implemented in 1998.

## Discussion

The government of South Africa has been successful in the extent to which it has made ART therapy available in the public sector and in negotiating drug prices. Currently, an estimated 2.1 million people are being kept alive on ART and the cost savings to the economy have been considerable. It is clear that expanding access will save more lives and, in the long run, will save money and eliminate HIV.

Universal access to early treatment is the only way to eliminate HIV in South Africa with currently available interventions.[3] The two most critical issues are drug supply and compliance. A regular and reliable supply of drugs must be assured. Stock-outs will create anger and mistrust among infected people and poor compliance, for any reason, will lead to viral rebound, treatment failure, ongoing transmission and drug resistance. These two considerations must be at the forefront of plans to effectively control and eventually eliminate HIV.

Other methods of support and control can play an important role.[3] To ensure high levels of compliance it will be necessary to deal with problems of stigma and discrimination and to ensure that there is strong community involvement and support for people living with HIV. Adult medical male circumcision will significantly reduce the prevalence of HIV among young men and have a secondary benefit for young women who are the group at greatest risk. Pre-exposure prophylaxis will provide additional protection for those that are unable to protect themselves as is often the case for women who believe that their partners may be infected but are unable to negotiate condom use. Condoms are highly effective if used properly and should be readily available to all that need them. Better control of other sexually transmitted infections is important in itself and will contribute further to the control of HIV. And while these interventions can make an important contribution to stopping the epidemic of HIV, universal access to early treatment provides an ideal entry point for each of them. Finally, by developing programmes that are firmly based in local communities it will be possible to provide training and education while employing community outreach workers thereby creating jobs and stimulating local economies.

It remains important to ensure the validity or otherwise of these results and to improve the model predictions. Here we have used the reported coverage of ART but this has never been directly measured. To do so all women who test positive for HIV in the annual ante-natal clinic surveys should be tested for the presence of anti-retroviral drugs, their viral loads should be measured and an incidence assay should be used to estimate incidence. We have also assumed a cost of US$250 per year for care and support of those infected with HIV. The results are sensitive to this assumption but this level of cost could be achieved if there is effective community support and mobilization. If this money were used to train and employ community health workers it would create jobs and provide an important stimulus to local economies in the poorest communities.

What will be needed is a big push as and when the Department of Health adopts the new WHO guidelines. Since an estimated 90% of all HIV positive people will then be eligible for immediate treatment it would be advisable to abandon the use of $CD4^+$ cell counts but to make viral load testing more widely available as a way of assessing the impact of the programme and of monitoring compliance. This analysis suggests that the rate at which people are currently being started on ART will have to be increased by about four times, but only for about two years. While this will require considerable organization it will only incur a marginal increase in costs for about five years and within a few years the savings in lives and money will be substantial.

What is needed now is informed leadership, a programme of universal testing and immediate access to treatment, combined with good operational research, while monitoring the impact and implementation of the new programme. The sooner that this is undertaken the more lives and money that will be saved. South Africa has the wherewithal to stop the epidemic; all that is needed is the commitment and determination to achieve an AIDS Free Generation.

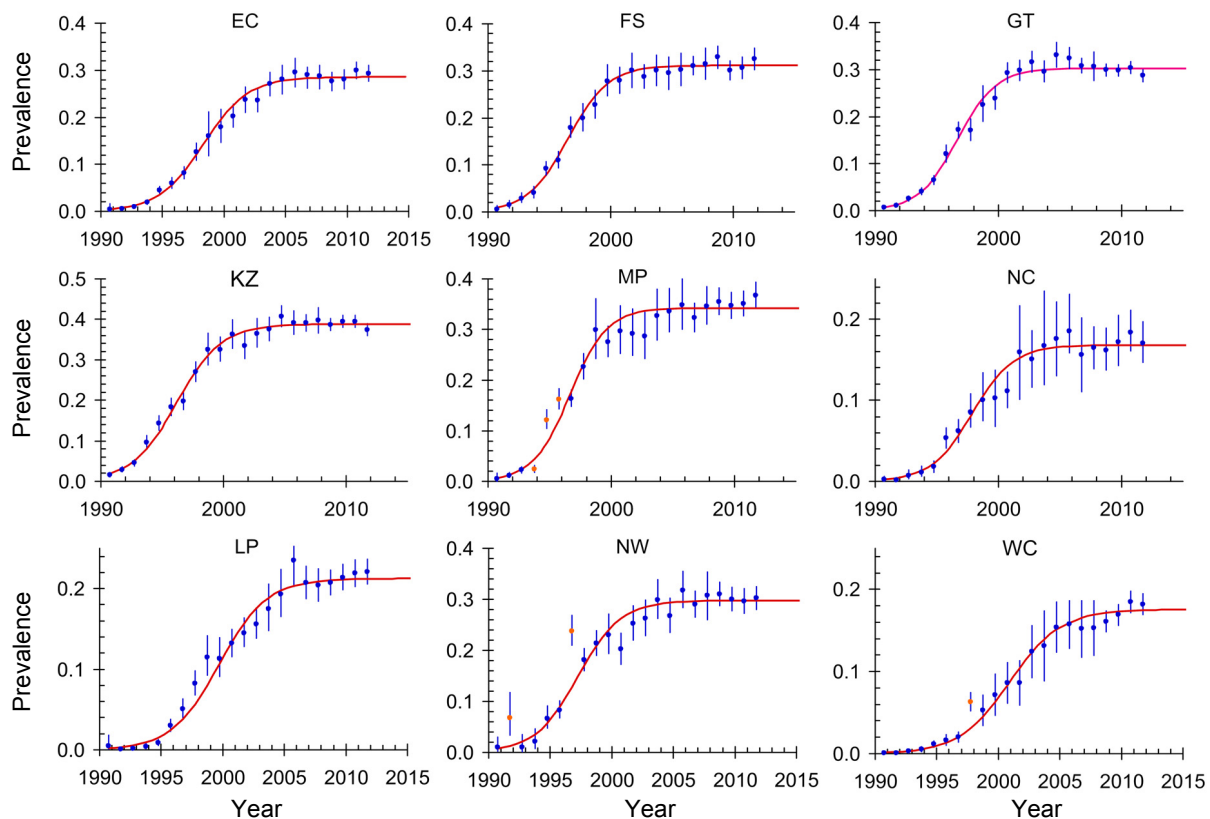

Figure 4. Prevalence of infection in women attending ante-natal clinics in South Africa. When fitting the data the points marked orange appear to be outlier and have not been included. EC: Eastern Cape; FS; Free State; GT: Gauteng; KZ: KwaZulu-Natal; MP: Mpumalanga; NC: Northern Cape; LP: Limpopo; NW: North West; WC: Western Cape.

## Appendix 1

The annual ante-natal clinic surveys in South Africa, started by Horst Küstner,[5-12] provide the best data on trends in the prevalence of HIV anywhere in the world. In all nine provinces in South Africa the prevalence of HIV has levelled off and is not declining with the possible exception of Gauteng (Figure 4). However, the timing and the extent of the epidemics varies among provinces. The epidemic in KwaZulu-Natal stabilized at 39% and reached half this value in March 1996. The epidemic in the Western Cape stabilized at 18% and only reached half this value in December 2000, almost four years later than in KwaZulu-Natal. The prevalence of HIV among women attending ante-natal clinics now appears to be stable in all provinces and, in the case of the Free State, has been stable for ten years.

The most plausible explanation for this is that the epidemic was introduced into in South Africa via trucking routes from Lusaka in Zambia to Durban in KwaZulu-Natal. In 1987 the prevalence of HIV among gold miners from Malawi was 4% but in men from all other countries, including South Africa, was less than 0.1%.[28] When this became known the mines stopped recruiting men from Malawi.[29] This could explain the observation that the epidemic started in the Free State and parts of Gauteng bordering on the Free State after KwaZulu-Natal but before the other provinces.

Table 5. Parameters for the fits to the provinces given in Figure 4. EC: Eastern Cape; FS: Free State; GT: Gauteng; KN: KwaZulu-Natal; LM: Limpopo; MP: Mpumalanga; NC: Northern Cape; NW: North West; WC: Western Cape

|    | Rate of increase/yr | Doubling time (yr) | Asymptote | Time to half-maximum (yr) |
|----|---------------------|--------------------|-----------|---------------------------|
| EC | 0.547 | 1.27 | 0.286 | 1998.34 |
| FS | 0.606 | 1.14 | 0.312 | 1996.54 |
| GT | 0.632 | 1.10 | 0.303 | 1996.68 |
| KN | 0.544 | 1.27 | 0.388 | 1996.22 |
| LM | 0.513 | 1.35 | 0.213 | 1999.73 |
| MP | 0.660 | 1.05 | 0.343 | 1996.65 |
| NC | 0.593 | 1.17 | 0.168 | 1997.90 |
| NW | 0.572 | 1.21 | 0.298 | 1997.26 |
| WC | 0.481 | 1.44 | 0.176 | 2000.97 |

It is probable that the main reason why the countries of southern Africa have the highest rates of HIV in the world is because of the system of oscillating migrant labour[30] which also provided the economic basis for Apartheid. Men are recruited from rural areas all over southern Africa and may spend much of their adult lives living without their wives or families. The rural areas experienced poverty and social disruption because of the lack of adult men and the men often had little to occupy them and to provide intimacy and social support apart from alcohol and sex.[31] This is supported by data for the Western Cape in 2001. Of the women attending public ante-natal clinics 62% were coloured and 37% were black. The overall prevalence of HIV was 8.8% but the prevalence of HIV among coloured women was 1.2% (0.7%–2.0%) and among black women was 21.6% (18.4%–25.3%). While conditions in the black townships may be worse than in the coloured townships the main difference is that the coloured townships are well established and have relatively stable populations while in the black townships there is a considerable and ongoing in and out migration from the Eastern Cape to the Western Cape.

## Appendix 2

The World Health Organization guidelines[25] of 2013 advise HIV-positive people to start ART if their $CD4^+$ cell count is less than 500/μL, if they have TB or Hepatitis B, if they are pregnant, under the age of five years, or in a sero-discordant relationship. Data on the distribution of $CD4^+$ cell counts in HIV-negative people[32] suggest that about 80% of all those currently infected with HIV in South Africa and not on ART will have a $CD4^+$ cell count below 500/μL. If we include the other groups of people that should start treatment irrespective of their $CD4^+$ cell count, then about 90% of all HIV positive people are currently eligible for ART.

If there is a need to triage people in order to treat those at greatest risk first, the sensible way to do this would be on the basis of each person's viral load. Individual $CD4^+$ cell counts can vary by an order of magnitude within populations,[32] the mean $CD4^+$ cell count can vary by a factor of two between populations,[32] and survival is independent of the initial $CD4^+$ cell count.[32,33] $CD4^+$ cell counts therefore have very little prognostic value except in that unfortunate circumstance when the count is very low by which time an infected person is likely to be in WHO clinical stages III or IV and in need of immediate treatment anyway. Considerable savings in time, human resources and money could be had by abandoning the use of $CD4^+$ cell counts for deciding on when to start ART. People with a high viral load, on the other hand, have a reduced life expectancy[34] and are more infectious than those with a low viral load.[35] It would therefore have the greatest benefit for individual patients and have the greatest impact on transmission if preference was given to people with high viral loads if the availability of anti-retroviral drugs is limited.[36]